\documentclass[tight,twocolumn]{aastex62}
\usepackage{apjfonts}
\usepackage{graphicx}
\usepackage{float}
\usepackage[caption=false]{subfig}
\usepackage{graphics} 
\usepackage{amssymb,amsmath}
\usepackage{color}

\def\lsim{\lower0.3em\hbox{$\,\buildrel <\over\sim\,$}}
\def\gsim{\lower0.3em\hbox{$\,\buildrel >\over\sim\,$}}

\newcommand{\msun}{\hbox{M$_{\odot}$}}

\begin{document}

\submitjournal{ApJ}

\title{Examining the Properties of Low-Luminosity Hosts of Type Ia Supernovae from ASAS-SN}
\shorttitle{ASAS-SN Low-L SN Ia Host Galaxies} 
\shortauthors{Holoien, et al.}

\author[0000-0001-9206-3460]{Thomas~W.-S.~Holoien}
\altaffiliation{NHFP Einstein Fellow}
\affiliation{The Observatories of the Carnegie Institution for Science, 813 Santa Barbara St., Pasadena, CA 91101, USA}

\author[0000-0002-7303-8144]{Vera~L.~Berger}
\affiliation{Department of Physics and Astronomy, Pomona College, 333 N. College Way, Claremont, CA 91711, USA}
\affiliation{The Observatories of the Carnegie Institution for Science, 813 Santa Barbara St., Pasadena, CA 91101, USA}

\author[0000-0001-9668-2920]{Jason~T.~Hinkle}
\affiliation{Institute for Astronomy, University of Hawai`i, 2680 Woodlawn Drive, Honolulu, HI 96822, USA}

\author[0000-0002-1296-6887]{L.~Galbany}
\affiliation{Institute of Space Sciences (ICE, CSIC), Campus UAB, Carrer de Can Magrans, s/n, E-08193 Barcelona, Spain}
\affiliation{Institut d’Estudis Espacials de Catalunya (IEEC), E-08034 Barcelona, Spain}

\author[0000-0001-6369-1636]{Allison~L.~Strom}
\affiliation{Department of Astrophysical Sciences, Princeton University, 4 Ivy Lave, Princeton, NJ 08544, USA}
\affiliation{Department of Physics and Astronomy and Center for Interdisciplinary Exploration and Research in Astrophysics (CIERA), Northwestern University, 2145 Sheridan Road, Evanston, IL 60208, USA}

\author{Patrick~J.~Vallely}
\affiliation{Department of Astronomy, The Ohio State University, 140 West 18th Avenue, Columbus, OH 43210, USA}

\author[0000-0003-0227-3451]{Joseph~P.~Anderson}
\affiliation{European Southern Observatory, Alonso de C\'ordova 3107, Casilla 19, Santiago, Chile}

\author{Konstantina~Boutsia}
\affiliation{Las Campanas Observatory, Carnegie Observatories, Casilla 601, La Serena, Chile}

\author[0000-0002-4235-7337]{K.~D.~French}  
\affiliation{Department of Astronomy, University of Illinois, 1002 W. Green St., Urbana, IL 61801, USA} 

\author{Christopher~S.~Kochanek}
\altaffiliation{Radcliffe Fellow}
\affiliation{Department of Astronomy, The Ohio State University, 140 West 18th Avenue, Columbus, OH 43210, USA}
\affiliation{Center for Cosmology and AstroParticle Physics (CCAPP), The Ohio State University, 191 W.\ Woodruff Ave., Columbus, OH 43210, USA}

\author{Hanindyo~Kuncarayakti}
\affiliation{Tuorla Observatory, Department of Physics and Astronomy, FI-20014 University of Turku, Finland}
\affiliation{Finnish Centre for Astronomy with ESO (FINCA), FI-20014 University of Turku, Finland}

\author[0000-0002-3464-0642]{Joseph~D.~Lyman}
\affiliation{Department of Physics, University of Warwick, Coventry, CV4 7AL, UK}

\author{Nidia~Morrell}
\affiliation{Las Campanas Observatory, Carnegie Observatories, Casilla 601, La Serena, Chile}

\author[0000-0003-0943-0026]{Jose~L.~Prieto}
\affiliation{N\'ucleo de Astronom\'ia de la Facultad de Ingenier\'ia y Ciencias, Universidad Diego Portales, Av. Ej\'ercito 441, Santiago, Chile}
\affiliation{Millennium Institute of Astrophysics, Santiago, Chile}

\author[0000-0001-6444-9307]{Sebasti\'{a}n F. Sánchez}
\affiliation{Instituto de Astronom\'{i}a, Universidad Nacional Aut\'{o}noma de Mexico, A. P. 70-264, C.P. 04510, M\'{e}xico, D.F., Mexico}

\author{K.~Z.~Stanek}
\affiliation{Department of Astronomy, The Ohio State University, 140 West 18th Avenue, Columbus, OH 43210, USA}
\affiliation{Center for Cosmology and AstroParticle Physics (CCAPP), The Ohio State University, 191 W.\ Woodruff Ave., Columbus, OH 43210, USA}

\author{Gregory~L.~Walth}
\affiliation{IPAC, California Institute of Technology, Mail Code 314-6, 1200 E. California Blvd., Pasadena, CA 91125}


\correspondingauthor{T.~W.-S.~Holoien}
\email{tholoien@carnegiescience.edu}

\date{\today}

\begin{abstract}
We present a spectroscopic analysis of 44 low-luminosity host galaxies of Type Ia supernovae (SNe Ia) detected by the All-Sky Automated Survey for Supernovae (ASAS-SN), using the emission lines to measure metallicities and star formation rates. We find that although the star formation activity of our sample is representative of general galaxies, there is some evidence that the lowest-mass SN Ia host galaxies (log($M_\star/M_\odot$)$<8$) in our sample have high metallicities compared to general galaxies of similar masses. We also identify a subset of 5 galaxies with particularly high metallicities. This highlights the need for spectroscopic analysis of more low-luminosity, low-mass SN Ia host galaxies to test the robustness of these conclusions and their potential impact on our understanding of SN Ia progenitors. 
\end{abstract}


\section{Introduction}
\label{sec:intro}

Type Ia supernovae (SNe Ia) are some of the most energetic events in the Universe, and due to empirical relations between their intrinsic brightness and other physical properties such as decline rate and color \citep[e.g.,][]{phillips93,hamuy96} they can be used as standardizeable candles to measure cosmological distances. SNe Ia were crucial to the discovery of the accelerating expansion of the Universe \citep{riess98,perlmutter99}. SNe Ia are also useful probes of physics under extreme conditions, are an end-stage of stellar evolution, and can significantly impact the evolution of their host galaxy environments \citep[e.g.,][]{maoz12,nomoto13,maoz17}. The study of these events thus impacts our understanding of a wide variety of astronomical sub-fields. Despite all this, the progenitor systems of SNe Ia are still not known, and their physics are not fully understood \citep[e.g.,][]{shappee17a,shappee18,shappee19,vallely19,tucker20}. There is also diversity in SNe Ia, with subtypes exhibiting different luminosities and decline rates that must be accounted for in order to measure their distances accurately \citep[e.g.,][]{phillips92,filippenko92a,filippenko92b,leibundgut93,foley13}.

One way to study the physical causes of the differences in SN Ia properties and the possible dependence of these differences on progenitor properties is to study the environments of SNe Ia. Previous studies of SNe Ia and their host galaxies have revealed correlations between the brightness, decline rate, and expansion rate of SNe Ia and the morphology, mass, and metallicity of their host galaxies \citep[e.g.,][]{filippenko89,branch93,hamuy00,gallagher05,gallagher08,howell09,lampeitl10,sullivan10}. This implies that one or more properties of SN Ia progenitors that are correlated with host galaxy properties are responsible for some of the observed diversity in SN Ia light curves.

Previous studies of the properties of SN Ia host galaxies have found that SN Ia hosts are generally consistent with the general population of galaxies. In particular, \citet{childress13} performed both a photometric and spectroscopic study of SN Ia hosts using the sample from the Nearby Supernova Factory \citep[SNFactory;][]{aldering02}. They found that SN Ia hosts had star formation activity and metallicities representative of normal galaxies, as had been found in previous work on photometric properties such as mass \citep[e.g.,][]{howell09,neill09}. These and earlier studies of SN Ia hosts largely focused on higher-mass, higher-luminosity host galaxies. This is likely a byproduct of earlier SN surveys being biased towards certain types of hosts due to observing strategy and/or survey design, and by a lack of follow-up resources to observe fainter, low-luminosity host galaxies in significant numbers.

In recent years, the proliferation large-area, rapid-cadence surveys such as the All-Sky Automated Survey for Supernovae \citep[ASAS-SN;][]{shappee14}, the Asteroid Terrestrial-impact Last Alert System \citep[ATLAS;][]{tonry18}, and the Zwicky Transient Facility \citep[ZTF;][]{bellm19} has resulted in large samples of SNe free from many of the host dependent biases of earlier SN surveys. This has allowed SN properties and rates to be correlated with host properties at much broader ranges than was previously possible. For example, \citet{brown19} were able to extend the observed trend in specific SN Ia rate with respect to host galaxy mass to roughly three orders of magnitude lower in mass than \citet{kistler13}. These modern samples, particularly nearby ones from bright-sky surveys such as ASAS-SN, are thus ideal for examining SN Ia host galaxies at low luminosities and masses. We can use these samples to test whether these hosts are as similar to the general population of galaxies, as has been seen with higher-mass and luminosity hosts. 

This is important to test for multiple reasons. First, if low-luminosity and low-mass SN Ia hosts differ from typical galaxies, this may affect the progenitors and properties of SNe Ia in these galaxies. Understanding how SNe Ia in these hosts may differ from those in more typical hosts will be important for developing a full physical understanding of SNe Ia and their explosion mechanisms. Second, if these low-luminosity hosts do not have properties similar to general galaxies, it would mean standard galaxy relations, e.g., the mass-metallicity relation, would not apply to this population of galaxies. Properties such as metallicity, which can only reliably be measured via observationally expensive spectroscopy, are often inferred based on photometrically measurable properties such as mass when studying SN Ia hosts \citep[e.g.,][]{howell09,neill09}. If low-luminosity and low-mass hosts deviate from standard relations, it is crucial to understand how they deviate so that we can correctly account for the differences when inferring properties such as metallicity in these galaxies.

This paper presents a spectroscopic study of the properties of 44 low-luminosity SN Ia host galaxies selected from the first three years of ASAS-SN. In Section~\ref{sec:sample} we discuss the galaxies in our sample and how they were selected, the spectroscopic dataset, how we measured emission line fluxes from the spectra, and how we translated these line fluxes into physical properties. In Section~\ref{sec:analysis} we analyze these properties and compare them to several samples of non-SN-host galaxies. Finally, in Section~\ref{sec:disc} we discuss our findings and the future directions of this work. Throughout this paper we assume a standard $\Lambda$CDM cosmology with H$_0=69.6$~km~s$^{-1}$~Mpc$^{-1}$, $\Omega_M=0.296$, and $\Omega_{\Lambda}=0.714$ \citep{wright06,bennett14}.


\section{Low-Luminosity SN Ia Host Galaxy Sample}
\label{sec:sample}

\subsection{Sample Details}
\label{sec:sample_details}

To select a representative sample of low-luminosity host galaxies of SNe Ia, we first started with the sample of SNe Ia used by \citet{brown19} to measure the relative specific SN Ia rate from the first 3 years of ASAS-SN. The ASAS-SN sample is ideal for a number of reasons: ASAS-SN surveys the entire sky systematically, meaning there is no bias towards a previously selected sample of galaxies or towards galaxies of a particular luminosity; ASAS-SN surveys the nearby Universe, meaning low-luminosity galaxies in the ASAS-SN sample should be close enough to observe spectroscopically; and the ASAS-SN sample was small enough for all possible SNe to be observed spectroscopically, lowering the likelihood of a SN being missed due to limited resources. 

From the \citet{brown19} sample we selected galaxies with $\log{L/L_\star}\leq-1.5$, calculating the luminosity of the galaxies based on the $K_S$-band magnitudes from \citet{holoien17b} and assuming $M_{\star,K_S}=-24.2$ \citep{kochanek01}. This resulted in a sample of 58 low-luminosity SN Ia host galaxies that we targeted for spectroscopic observation. The final sample presented here comprises 44 of these galaxies for which we were able to obtain spectra with high enough signal-to-noise to measure the emission line fluxes needed for our analyses. These are primarily H$\alpha$, H$\beta$, the [\ion{O}{3}] $\lambda4959/5007$ doublet, and the [\ion{N}{2}] $\lambda6548/6583$ doublet, though we also measure several others when possible. Our reduced spectra are available in the online journal. For completeness, we also include in our dataset the spectra of nine additional galaxies which were observed as part of our observing program, but did not yield well-measured emission line fluxes.

The telescopes and instruments used to obtain spectra for our sample were: (1) the Inamori-Magellan Areal Camera and Spectrograph \citep[IMACS;][]{dressler11} on the 6.5-m Magellan-Baade telescope; (2) the Low-Dispersion Survey Spectrograph 3 (LDSS-3) on the 6.5-m Magellan Clay telescope; (3) the Multi-Object Double Spectrograph \citep[MODS;][]{Pogge2010} on the dual 8.4~m Large Binocular Telescope (LBT); (4) the Multi-Unit Spectroscopic Explorer \citep[MUSE;][]{bacon14}, located at the Nasmyth B focus of Yepun, the VLT UT4 telescope at Cerro Paranal Observatory; and (5) the Potsdam Multi Aperture Spectograph \citep[PMAS;][]{roth05} mounted on the 3.5 m telescope of the Centro Astronomico Hispano-Aleman (CAHA) at the Calar Alto Observatory. With the exception of the MUSE and PMAS spectra, these spectra consisted of long-slit spectra obtained at the parallactic angle and centered on the host nucleus, with total integration times typically between 1 and 4 hours. 

We used {\sc Iraf} to reduce our IMACS and LDSS-3 spectra following standard procedures, including bias subtraction, flat-fielding, one-dimensional spectral extraction, and wavelength calibration using a comparison lamp spectrum. The MODS spectra were reduced using the MODS spectroscopic pipeline\footnote{\url{http://www.astronomy.ohio-state.edu/MODS/Software/modsIDL/}}. We flux-calibrated our spectra using observations of spectrophotometric standard stars obtained on the same nights as our galaxy spectra.

Seven of the spectra in our sample were obtained with MUSE. Its modular structure is composed of 24 identical integral field unit (IFU) modules that together sample, in Wide Field Mode (WFM), a near-contiguous 1 arcmin$^2$ FoV with spaxels of 0.2 $\times$ 0.2 arcsec, a wavelength coverage of 4650-9300 \AA~and a mean resolution of R $\sim$3000. This produces $\sim$100,000 spectra per pointing. These observations were obtained by the All-weather MUSE Supernova Integral-field of Nearby Galaxies (AMUSING\footnote{\href{https://amusing-muse.github.io/}{https://amusing-muse.github.io/}}; \citealt{galbany16a}; Galbany et al. in prep.) survey. This survey has been running for 10 semesters, and has compiled observations for more than 600 nearby SN host galaxies.

One additional spectrum (UGC 08503) was obtained with PMAS in PPak mode \citep{verheijen04,kelz06}. PPak consists of a fiber bundle of 382 fibers with 2.7" diameter, 331 of which are ordered in a single hexagonal bundle, and the remaining fibers are used for sky measurements and calibration purposes. Observations were performed using the V500 grating, which has a spectral resolution of $\sim$6 \AA~in the wavelength range 3750$-$7300 \AA. The final product is a 3D data cubes with a 100\% covering factor within a hexagonal FoV of $\sim$1.3 arcmin$^2$ with 1"$\times$1" pixels, which correspond to $\sim$4000 spectra per object. This observation is part of the PMAS/PPak Integral field Supernova hosts COmpilation (PISCO\footnote{\href{https://github.com/lgalbany/pisco}{https://github.com/lgalbany/pisco}}; \citealt{galbany18}), a project that aimed at building a sample of supernova host galaxies for environmental studies \citep{galbany14,galbany16a}. As of May 2022, the PISCO sample contained 363 galaxies.

To extract the global spectra of the host galaxies from the MUSE and PMAS data, we defined an elliptical aperture by fitting an elliptical S\'ersic profile to the galaxy light using an image obtained by compressing the cube in the wavelength direction, simulating an image with a flat filter from $\sim$4800 to $\sim$9300 \AA. Circular apertures were placed at the position of foreground stars selected from the Gaia EDR3 catalogue \citep{gaiaedr3}, and the flux within the apertures was removed and interpolated from the nearby pixels outside the aperture.

\subsection{Line Measurements}
\label{sec:line_measurements}

\begin{figure*}
\centering
\includegraphics[width=0.99\textwidth]{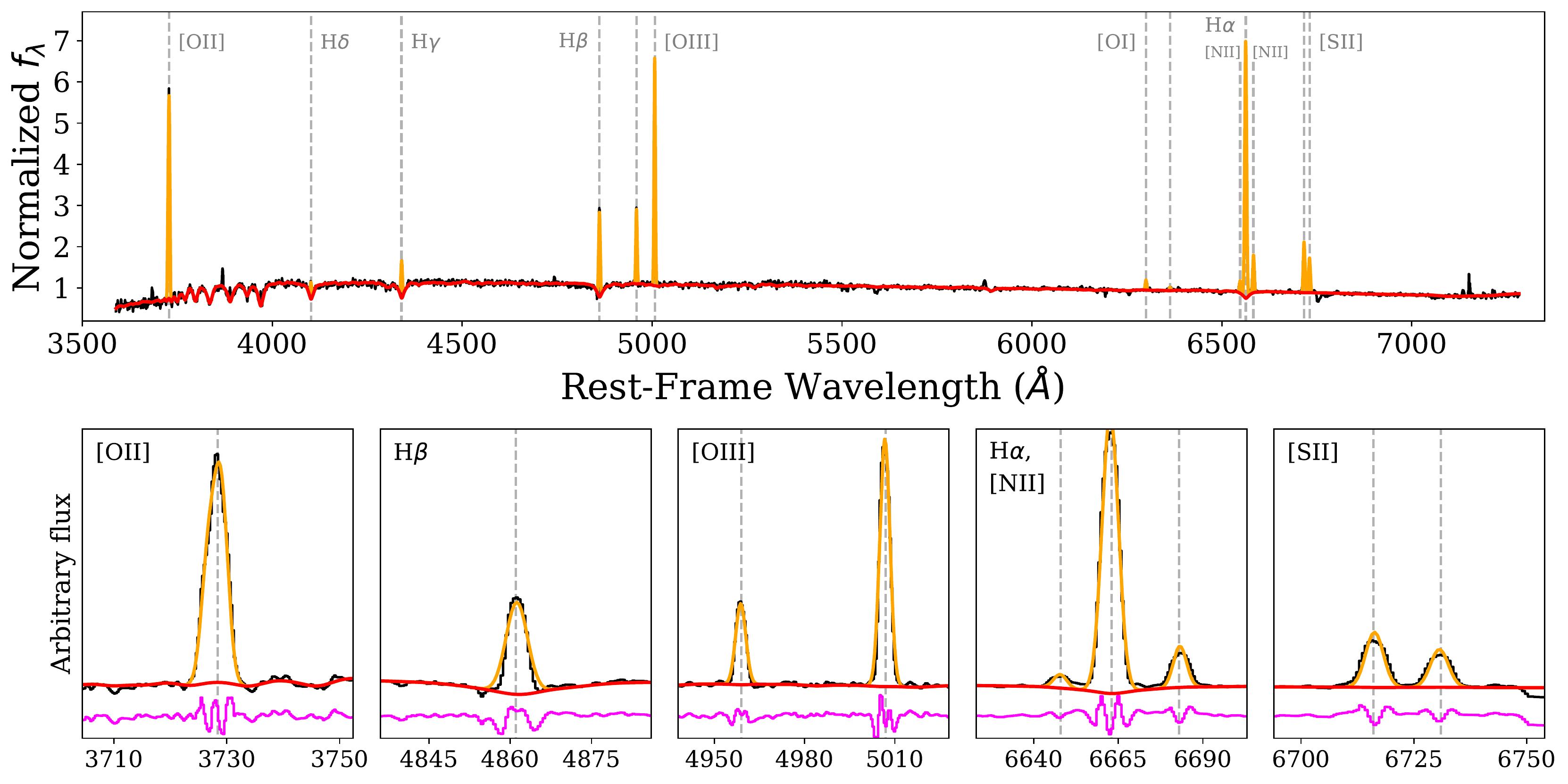}
\caption{The MODS/LBT of the galaxy 2MASX J03051061+3754003, the host of the supernova ASASSN-15li, and fits from {\sc pPXF}. The black curve is the observed spectrum, the red curve is the stellar continuum fit, and the gold curve is the combined stellar continuum$+$emission line fit. The lines included in the fit are marked and labelled in grey. The lower panels show zoomed-in views of several fitted lines as well as the residual difference between the data and the combined model in magenta. The labels in the upper-left corners give the lines being shown in each panel.}
\label{fig:fit_example}
\end{figure*}

In order to calculate the metallicities and star formation rates of the galaxies in our sample, we first measured the fluxes of the emission lines in the spectra. To do so, we used the Python code {\sc pPXF} \citep{cappellari12,cappellari17} to simultaneously fit the stellar background needed to model the effects of stellar absorption lines and the emission line fluxes. {\sc pPXF} uses the Penalized Pixel Fitting method \citep{cappellari04,cappellari17} to extract the galaxy stellar and gas kinematics, stellar population, and gas emission lines by fitting a set of templates to an observed spectrum. After fitting a stellar template to the observed spectrum, {\sc pPXF} subtracts the template from the spectrum and fits Gaussian profiles to each of the emission lines. Each line is fit individually, with the exceptions of the [\ion{O}{3}] $\lambda4959/5007$, [\ion{O}{1}] $\lambda6300/6364$, and [\ion{N}{2}] $\lambda6548/6583$ doublets, where the brighter of the two lines is fit and the lines are assumed to have a 3:1 flux ratio. 

We obtained line fluxes by integrating the Gaussian model of each emission component. We calculated equivalent widths for H$\alpha$ and H$\beta$ by estimating the continuum using regions near the lines in the observed spectrum (prior to the {\sc pPXF} template subtraction) and dividing the measured line flux by the continuum estimate. An example host galaxy spectrum, the best-fit stellar template, and the combined stellar template$+$emission lines model is shown in Figure~\ref{fig:fit_example}. The {\sc pPXF} emission line models are shown in gray.

Finally, we estimated dust extinction in each galaxy by comparing the Balmer decrement ($H\alpha / H\beta$) to the expected value of 2.86 for Case B recombination \citep{osterbrock89}. Roughly 29\% of the galaxies in our sample have Balmer decrements consistent with the Case B ratio within $3\sigma$. We assume zero extinction for the three galaxies in our sample with $H\alpha / H\beta < 2.86$. We then calculate $E(B-V)$ and correct the flux of each spectroscopic line for extinction using a \citet{cardelli89} extinction law with $R_V=3.1$.  Our final extinction-corrected line fluxes and equivalent widths are presented in Table~\ref{tab:line_fluxes}.

\begin{deluxetable*}{lcccc}
\tabletypesize{\footnotesize}
\tablecaption{Emission line fluxes of our low-L SN Ia host galaxy sample}
\tablehead{
\colhead{} &
\multicolumn{4}{c}{Galaxy Name$^{1}$} \vspace{-6pt} \\
\colhead{Line} &
\colhead{2MASX J03051061+3754003} &
\colhead{AGC 331536} &
\colhead{APMUKS(BJ) B032028.93-441621.0} &
\colhead{CGMW 2-2125}}
\startdata
\textrm{[OII]} $\lambda 3727+3729$ & $5748.0\pm212.4$ & $78.63\pm6.53$ & --- & --- \\ 
H$\delta$ & $98.07\pm12.1$ & $4.49\pm0.49$ & $156.7\pm9.68$ & $29.7\pm18.3$ \\ 
H$\gamma$ & $489.7\pm19.28$ & $8.54\pm0.72$ & $309.2\pm13.92$ & $30.64\pm17.45$ \\ 
H$\beta$ & $1263.0\pm36.57$ & $21.65\pm1.4$ & $714.9\pm24.89$ & $71.91\pm30.22$ \\ 
\textrm{[OIII]} $\lambda 4959$ & $1071.0\pm29.22$ & $14.44\pm0.89$ & $635.8\pm20.89$ & $32.3\pm12.98$ \\ 
\textrm{[OIII]} $\lambda 5007$ & $3170.0\pm85.39$ & $43.15\pm2.62$ & $1889.0\pm61.3$ & $95.02\pm37.71$ \\ 
\textrm{[OI]} $\lambda 6300$ & $124.9\pm8.1$ & $2.27\pm0.45$ & $24.09\pm6.8$ & $11.46\pm7.16$ \\ 
\textrm{[OI]} $\lambda 6364$ & $41.2\pm2.66$ & $0.75\pm0.15$ & $7.98\pm2.24$ & $3.78\pm2.34$ \\ 
\textrm{[NII]} $\lambda 6548$ & $142.5\pm3.78$ & $2.85\pm0.19$ & $84.94\pm3.04$ & $17.06\pm5.27$ \\ 
H$\alpha$ & $3608.0\pm71.15$ & $61.89\pm2.75$ & $2043.0\pm48.56$ & $205.3\pm58.61$ \\ 
\textrm{[NII]} $\lambda 6583$ & $425.8\pm11.24$ & $8.52\pm0.58$ & $253.7\pm9.03$ & $50.82\pm15.6$ \\ 
\textrm{[SII]} $\lambda 6716$ & $567.2\pm13.39$ & $10.39\pm0.65$ & $323.3\pm10.34$ & $92.75\pm26.35$ \\ 
\textrm{[SII]} $\lambda 6731$ & $388.0\pm10.82$ & $9.2\pm0.62$ & $218.9\pm8.77$ & $21.98\pm8.77$ \\ 
\hline
H$\alpha$ EW (\AA) & $41.07\pm0.1$ & $19.18\pm0.13$ & $96.76\pm0.31$ & $11.32\pm0.35$ \\ 
H$\beta$ EW (\AA) & $8.51\pm0.07$ & $4.71\pm0.08$ & $16.74\pm0.16$ & $2.32\pm0.27$ \\ 
\enddata 
\tablenotetext{1}{For cases where a galaxy has not been previously catalogued by any survey, and thus does not have an archival name or designation, we use ``Uncatalogued'' for the name and give the corresponding SN name in parentheses. \vspace{-5pt}}
\tablecomments{Extinction-corrected line fluxes and equivalent widths for the galaxies in our sample. Line fluxes were calculated using {\sc pPXF} and corrected for extinction using the Balmer decrement, as described in Section~\ref{sec:line_measurements}. All fluxes are given in units of $10^{-17}$~erg~s$^{-1}$~cm$^{-2}$ and equivalent widths are given for H$\alpha$ and H$\beta$ in \AA. A portion of the table is shown here for guidance regarding its form and content; the full table is available in machine-readable format in the online journal.}
\label{tab:line_fluxes}
\end{deluxetable*}

\begin{figure*}
\centering
\includegraphics[width=0.99\linewidth]{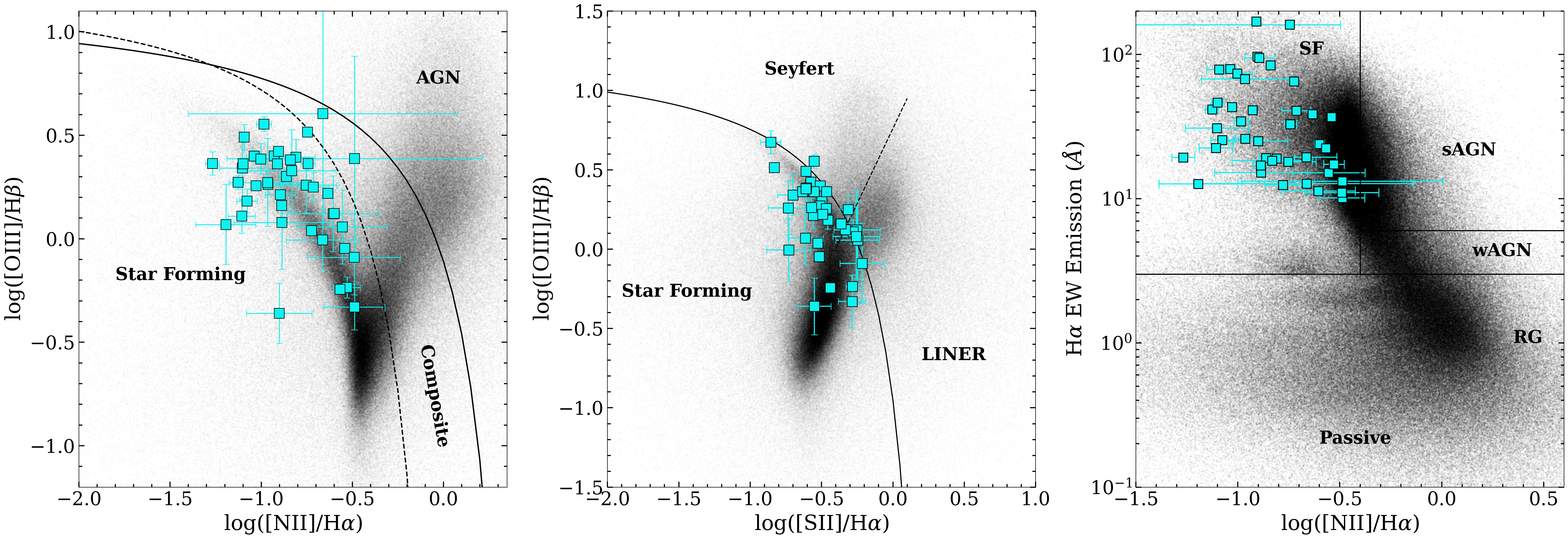}
\caption{Emission line diagnostic diagrams for our sample (blue squares). \textit{Left Panel}: BPT diagram showing log([\ion{O}{3}] / H$\beta$) vs. log([\ion{N}{2}] / H$\alpha$) \citep{baldwin81, veilleux87}. The solid line shows the separation between AGNs (above) and \ion{H}{2}/star-forming regions (below) based on theoretical models \citep{kewley01}, while the dotted line shows the same classifications based on empirical models \citep{kauffmann03}. \textit{Center Panel}: log([\ion{O}{3}] / H$\beta$) vs. log([\ion{S}{2}] / H$\alpha$) diagnostic from \citet{veilleux87}. The solid line separates AGNs from \ion{H}{2} regions as in the BPT diagram \citep{kewley01}, and the diagonal dashed line separates Seyferts (above) and LINERs (below) \citep{kewley06}. \textit{Right Panel}: H$\alpha$ emission line equivalent width (EW$_{H\alpha}$) compared to log([\ion{N}{2}] / H$\alpha$), also called the WHAN diagram \citep{cidfernandes11}. Lines separating star-forming galaxies (SF), strong AGNs (sAGN), weak AGNs (wAGN), and passive/``retired galaxies'' (RG) are shown \citep{cidfernandes11}. In all panels, galaxies from SDSS DR8 \citep{eisenstein11} are shown in black.}
\label{fig:line_diagnostics}
\end{figure*}

In Figure~\ref{fig:line_diagnostics} we show several line diagnostic diagrams with our low-luminosity host sample compared to the Sloan Digital Sky Survey Data Release 8 \citep[SDSS DR8;][]{eisenstein11} sample of galaxies. The left panel shows the BPT diagram \citep{baldwin81}, the center panel shows the log([\ion{S}{2}]/H$\alpha$) vs. log([\ion{O}{2}]/H$\beta$) diagram \citep{veilleux87}, and the right panel shows the WHAN (H$\alpha$ equivalent width vs. [\ion{N}{2}]/H$\alpha$) diagram \citep{cidfernandes11}. Lines dividing different types of galaxies (typically star-forming vs. AGN) are shown in each panel and described in the figure caption. In all three cases, the vast majority of the galaxies in our sample clearly lie in the star-forming or \ion{H}{2} regions of the diagrams, though we do note that a handful are also consistent with falling in the ``Composite'' region of the BPT diagram or the Seyfert region of the [\ion{S}{2}]/H$\alpha$ diagram. Our sample is largely clustered in a region of the BPT diagram that is populated by low-mass star-forming galaxies, which reinforces the low-mass nature of our sample (see Section~\ref{sec:phys_props}). Given the strong emission lines present in our spectra and the low-luminosity, low-mass nature of the sample, their locations in these diagnostic figures is unsurprising.

\subsection{Physical properties}
\label{sec:phys_props}

The analyses presented here primarily consider four physical properties of our galaxies: the stellar mass, the metallicity, the star formation rate, and the specific star formation rate. Stellar masses for the galaxies were calculated by \citet{brown19}. They computed masses for the galaxies using the Fitting and Assessment of Synthetic Templates \citep[\textsc{fast};~][]{kriek09} to fit spectral energy distributions to archival photometry of the galaxies. The archival photometry included GALEX $NUV$, SDSS or Pan-STARRS optical, 2MASS $JHK_S$, and WISE $W1$ and $W2$ data, or a subset thereof when reliable magnitudes were not available for all bands. Their fits assumed a \citet{cardelli89} extinction law with $R_V=3.1$ and Galactic extinction taken from \citet{schlafly11}, and used an exponentially declining star-formation history, a \citet{salpeter55} initial mass function (IMF), and the \citet{bruzual03} stellar population models for the fits. Due to heterogenous photometry used for the fit, they assumed a minimum uncertainty of 0.1~mag for all magnitudes used to fit the SEDs in order to avoid the fit being biased by artificially small uncertainties in a given filter. A comparison of their results with the values from the MPA-JHU Galspec pipeline \citep{kauffmann03} indicated that the masses they derive are largely consistent with those from Galspec, and we adopt the \citet{brown19} masses in our analyses.

One galaxy in our sample, the uncatalogued host of CSS141227:132741-111252, was not included in the \citet{brown19} analysis, as it was previously not detected in archival data. We obtained $grizy$ Kron magnitudes of the host from the Panoramic Survey Telescope and Rapid Response System \citep[Pan-STAARS;][]{chambers16,flewelling16} and infrared $W1$ and $W2$ magnitudes from the {\it Wide-field Infrared Survey Explorer} \citep[{\it WISE};][]{wright10,allwiseDOI} AllWISE catalog and fit the SED using \textsc{fast} in the same manner as \citet{brown19}. We find that the host has a stellar mass of $M_\star=(1.35^{+0.06}_{-0.03})\times10^8$~\msun, consistent with a dwarf galaxy and similar to many galaxies in our sample. One additional galaxy in our sample, GALEXASC J134316.80-313318.2, has limited archival data available with which to fit the SED, and only has an upper limit on its mass presented in \citet{brown19}. We use the same limit for the mass here.

Translating emission line fluxes into gas-phase metallicities is highly dependent on the choice of calibration, with different calibrations often disagreeing by as much as $\sim0.5$~dex \citep[e.g.,][]{kewley08}. Further, metallicity calibrations are only valid for specific ranges of line ratios or metallicities and/or based on different estimates of physical properties that correlate with metallicity. Finding a single method that can be used for a wide range of metallicities is difficult. Empircial metallicity calibrations are generally preferred as they are based on the electron temperature abundance scale, but they are typically calculated based on samples that do not cover the full range of galaxy parameter space. 

We calculate gas-phase metallicities for the galaxies in our sample using the empirical calibrations derived by \citet{curti17}. \citet{curti17} calculated a new set of empirical metallicity calibrations using a uniform application of the $T_e$ method across the full metallicity range spanned by SDSS galaxies. Their calibrations span more than 1 dex in metallicity, have been calculated for several commonly used line ratios, and the calibrations for different line ratios are consistent with one another to within 0.05 dex. They are ideal for both our low-luminosity SN host galaxy sample and the comparison samples we use (see Section~\ref{sec:analysis} below). 

We use the \citet{curti17} N2 method, based on the log([\ion{N}{2}]~$\lambda6583$/H$\alpha$) ratio, to calculate the metallicities presented in our analyses. For cases where the uncertainties on the log([\ion{N}{2}]/H$\alpha$) ratio were larger than the ratios themselves, we instead calculated 1$\sigma$ upper limits on the line ratio and then calculated 1$\sigma$ upper limits on the metallicities from these. One galaxy in our sample, GALEXASC J005328.92-362154.1, does not have a clear detection of [\ion{N}{2}]~$\lambda6583$ due to being blended with a very bright H$\alpha$ line, and we exclude this galaxy from analysis involving metallicity below.

\begin{deluxetable*}{ccccccccc}
\tabletypesize{\footnotesize}
\tablecaption{Physical properties of our low-L SN Ia host galaxy sample}
\label{tab:gal_props} 
\tablehead{
\colhead{} &
\colhead{} &
\colhead{} &
\colhead{Distance} &
\colhead{$\log{(M_{\star})}$} &
\colhead{} &
\colhead{$\log{(\textrm{SFR})}$} &
\colhead{$\log{(\textrm{sSFR})}$} &
\colhead{} \vspace{-6pt} \\
\colhead{Galaxy Name$^{1}$} &
\colhead{SN Name} &
\colhead{Redshift$^{2}$} &
\colhead{(Mpc)} &
\colhead{($M_{\odot}$)} &
\colhead{$12+\log{(O/H)}^{3}$} &
\colhead{$M_{\odot}~\textrm{yr}^{-1}$} &
\colhead{Gyr$^{-1}$} &
\colhead{$M_{K_S}^{4}$}}
\startdata
2MASX J03051061 & ASASSN-15ti & 0.0172 & 75.1 & $8.72_{-0.04}^{+0.04}$ & $8.45_{-0.01}^{+0.01}$ & $-0.87_{-0.01}^{+0.01}$ & $-0.59_{-0.05}^{+0.05}$ & $-18.18\pm0.20$ \\ 
AGC 331536 & ASASSN-16ke & 0.028 & 123.0 & $8.60_{-0.11}^{+0.28}$ & $8.49_{-0.02}^{+0.02}$ & $-2.21_{-0.02}^{+0.02}$ & $-1.81_{-0.30}^{+0.13}$ & $-17.26\pm0.11$ \\ 
APMUKS(BJ) B032028.93 & ASASSN-16dx & 0.0295 & 130.0 & $8.05_{-0.17}^{+0.60}$ & $8.47_{-0.01}^{+0.01}$ & $-0.64_{-0.01}^{+0.01}$ & $0.310_{-0.61}^{+0.18}$ & $-18.37\pm0.20$ \\ 
CGMW 2-2125 & ASASSN-16jq & 0.018 & 78.6 & $9.10_{-0.16}^{+0.36}$ & $8.62_{-0.10}^{+0.12}$ & $-2.08_{-0.15}^{+0.11}$ & $-2.18_{-0.51}^{+0.27}$ & $>-18.88$ \\ 
ESO 113-G047 & ASASSN-14me & 0.018 & 78.6 & $8.46_{-1.13}^{+0.62}$ & $8.35_{-0.02}^{+0.02}$ & $-0.54_{-0.01}^{+0.01}$ & $-0.00_{-0.63}^{+1.14}$ & $-17.19\pm0.11$ \\ 
ESO 357-G005 & ASASSN-15pz & 0.0149 & 64.9 & $7.28_{-0.34}^{+0.86}$ & $8.65_{-0.10}^{+0.13}$ & $-2.34_{-0.14}^{+0.11}$ & $-0.62_{-1.00}^{+0.45}$ & $-16.67\pm0.11$ \\ 
ESO 466-G032 & 2016ekg & 0.0171 & 74.6 & $8.79_{-0.05}^{+0.03}$ & $8.56_{-0.01}^{+0.01}$ & $-2.15_{-0.01}^{+0.01}$ & $-1.94_{-0.04}^{+0.06}$ & $-18.16\pm0.20$ \\ 
ESO 509-IG064 & ASASSN-16hp & 0.008663 & 37.6 & $8.57_{-0.05}^{+0.01}$ & $8.46_{-0.00}^{+0.00}$ & $-1.49_{-0.01}^{+0.01}$ & $-1.06_{-0.02}^{+0.06}$ & $-17.88\pm0.10$ \\ 
GALEXASC J000802.78 & ASASSN-15rq & 0.0236 & 104.0 & $8.43_{-0.67}^{+0.70}$ & $8.47_{-0.04}^{+0.04}$ & $-1.29_{-0.04}^{+0.04}$ & $-0.72_{-0.74}^{+0.71}$ & $-17.38\pm0.11$ \\ 
\enddata 
\tablenotetext{1}{For cases where a galaxy has not been previously catalogued by any survey, and thus does not have an archival name or designation, we use ``Uncatalogued'' for the name. Some galaxy names are abbreviated here to prevent the width of the table from extending beyond the page, but full names are given in the version of the table available in the online journal. \vspace{-5pt}}
\tablenotetext{2}{Redshifts that had not been previously measured or that differ substantially from archival measurements are denoted with an asterisk. \vspace{-5pt}}
\tablenotetext{3}{Metallicities are calculated using the N2 method of \citet{curti17}. \vspace{-5pt}}
\tablenotetext{4}{$K_S$-band magnitudes are taken from the 2MASS catalogs when available, or calculated from the WISE $W1$-band magnitudes when not. If the galaxy is not detected in either catalog, we assume an upper limit of $m_{K_S}>15.6$, as described in the text. \vspace{-5pt}}
\tablecomments{Physical parameters for the galaxies in our sample. Descriptions of how each property was calculated are given in Section~\ref{sec:phys_props}. A portion of the table is shown here for guidance regarding its form and content; the full table is available in machine-readable format in the online journal.}
\vspace{-0.8cm}
\end{deluxetable*}

The current star formation rate (SFR) is typically calculated using the H$\alpha$ line luminosity. As part of the line measuring process, we also measured the redshifts of the galaxies in our sample, or verified the archival redshifts that had previously been reported. For each galaxy we used our spectroscopic redshift to calculate the luminosity distance, and then calculated the total rest-frame H$\alpha$ luminosity from the reddening-corrected H$\alpha$ line flux. We converted this luminosity to an SFR using the \citet{kennicutt98} formula and multiply by 0.7 to convert from a \citet{salpeter55} IMF to a \citet{chabrier03} IMF. After calculating the SFRs of our sample, we also calculate the specific star formation rates (sSFRs) by dividing the SFRs by the stellar masses from the broadband SED fits. For the case of GALEXASC J134316.80-313318.2 with only an upper limit on its mass, we treat the calculated sSFR as a lower limit.

Though we do not examine luminosity in detail in our analyses, our sample was selected based on a luminosity threshold, and we also report the luminosities of our galaxies based on Two Micron All-Sky Survey \citep[2MASS;][]{skrutskie06,2massDOI} $K_S$-band magnitudes. We collected the $K_S$-band magnitudes for the galaxies in our sample from the ASAS-SN Bright Supernova Catalogs \citep{holoien16d,holoien17a,holoien17b} from which our sample was drawn. As the $K_S$-band magnitude has a well-defined luminosity function \citep{kochanek01}, it provides a natural estimate for the stellar luminosities of the galaxies in our sample. In the ASAS-SN catalogs, the authors obtained the 2MASS magnitudes from the 2MASS Extended Source Catalog when available, and from the 2MASS Point Source Catalog when the galaxies were not detected in the Extended Source Catalog. For galaxies not detected in 2MASS but detected in WISE $W1$ data, they estimated the $K_S$-band magnitude by subtracting the average $K_s - W1$ offset of $-0.51$ mag calculated from all galaxies detected in both filters. Finally, for galaxies not detected by either 2MASS or WISE, they assumed an upper limit of $K_s > 15.6$, equal to the faintest detected galaxy in their sample. After collecting the $K_S$ apparent magnitudes from the ASAS-SN catalogs, we then converted to absolute magnitudes using the distance moduli calculated from the galaxies' spectroscopic redshifts.

The coordinates, redshifts, distances, masses, metallicities, SFRs, sSFRs, and absolute $K_S$-band magnitudes of the galaxies in our sample are presented in Table~\ref{tab:gal_props}. Redshifts that had not been previously measured or that differ from previously reported measurements are noted with an asterisk.

As our sample largely consists of longslit spectra centered on the galaxy nuclei, there is a possibility that the properties we derive from these spectra are not truly ``global''. Because of this, we performed some checks to verify that the line fluxes and galaxy properties we derive can be robustly compared to other galaxy samples. First, looking at the acquisition images of our spectra, the majority of our sample consists of small galaxies whose light is mostly contained within the slit. Thus, only for a handful of cases is there a chance of significant slit losses. Further, a handful of our sample were observed multiple times, either with different longslit orientations or with both MUSE and one of our longslit instruments. We find that the properties we derive for the galaxies with multiple observations are consistent between observations, implying that we are able to measure galaxy properties robustly with our longslit spectra. We also compared our derived SFRs to the SFRs derived by \citet{brown19} from their photometric SED fits, finding that the SFRs we derive from the H$\alpha$ luminosities are consistent with their SFRs. Finally, we take steps to ensure that our comparison samples are capturing a similar fraction of host light as our longslit spectra so that they will be directly comparable to our sample (see Section~\ref{sec:analysis}).


\section{Analysis and Comparison to other Galaxy Samples}
\label{sec:analysis}

To determine how our SN host galaxies compare to general galaxies, we compare our sample to several samples from literature. Our primary comparison sample is the SDSS DR8 sample \citep{eisenstein11}, for which line flux measurements and galaxy property measurements are available in the MPA-JHU value added catalogs. Because the SDSS catalog values are based on spectra taken with fibers of $2-3$ arcsecond width, their galaxy properties are not truly ``global'' properties, as any light outside of the fiber width would not be included. To make their sample more directly comparable to ours, we used only those SDSS galaxies with redshifts of $z<0.05$, the same redshift range of our sample, so that the area covered by their fibers would be roughly comparable to the area covered by the longslits we used for our spectra. We also only include galaxies with $\log{(M_\star/M_\odot)}\geq7$, as below this value the line fluxes and derived metallicities are not reliable.

As the SDSS sample is primarily concentrated toward higher masses and luminosities than our low-luminosity sample, we also selected two low-luminosity samples of galaxies for comparison. The first, compiled by \citet[][hereafter B12]{berg12}, is consists of longslit spectroscopic observations of \ion{H}{2} regions from 42 low-luminosity galaxies in the Spitzer Local Volume Legacy (LVL) survey. The second, presented in \citet[][hereafter H18]{hsyu18}, consists of longslit spectroscopic observations of 45 low-metallicity blue compact dwarf galaxies selected from SDSS Data Release 12 \citep{alam15}. Although the latter sample was selected for low metallicity rather than luminosity, these systems are also low-mass and low-luminosity. As both samples use longslit spectra with similar widths as our data, we do not perform any additional corrections to make them comparable to our sample. 

For the SDSS sample, we used the stellar masses and SFRs from the MPA-JHU catalog and calculated sSFRs using those. As the MPA-JHU catalogs used a \citet{kroupa01} IMF to calculate masses and SFRs, we converted their measurements to a \citet{salpeter55} IMF by multiplying them by a factor of 1.5 \citep{brinchmann04}, then converted the SFRs to a \citet{chabrier03} IMF by multiplying by 0.7 so that they would be directly comparable to the other samples. For the B12 sample, we obtained masses from their paper. As their paper provided spectra of \ion{H}{2} regions, rather than of the galaxies as a whole, we were unable to use the reported H$\alpha$ fluxes to calculate H$\alpha$ luminosities and SFRs for the galaxies. Instead, we use the H$\alpha$ luminosities for the LVL galaxies from \citet{kennicutt08} to calculate the SFRs and sSFRs of the B12 sample\footnote{Four of the galaxies in the \citet{berg12} sample do not have H$\alpha$ luminosities in \citet{kennicutt08}, and we exclude these from our analyses of SFR and sSFR.}. For galaxies in the B12 sample with multiple spectra of \ion{H}{2} regions, we use the same SFRs and masses and calculate the metallicities for each \ion{H}{2} region separately. Finally, for the H18 sample, we obtained masses and SFRs from their paper.

For all three samples, we used [NII]~$\lambda6583$ and H$\alpha$ fluxes to calculate the Curti N2 metallicities as we did for our sample. We calculated upper limits on the metallicities when the [NII]~$\lambda6583$ and H$\alpha$ lines were not robustly detected.

\begin{figure}
\centering
\includegraphics[width=0.99\linewidth]{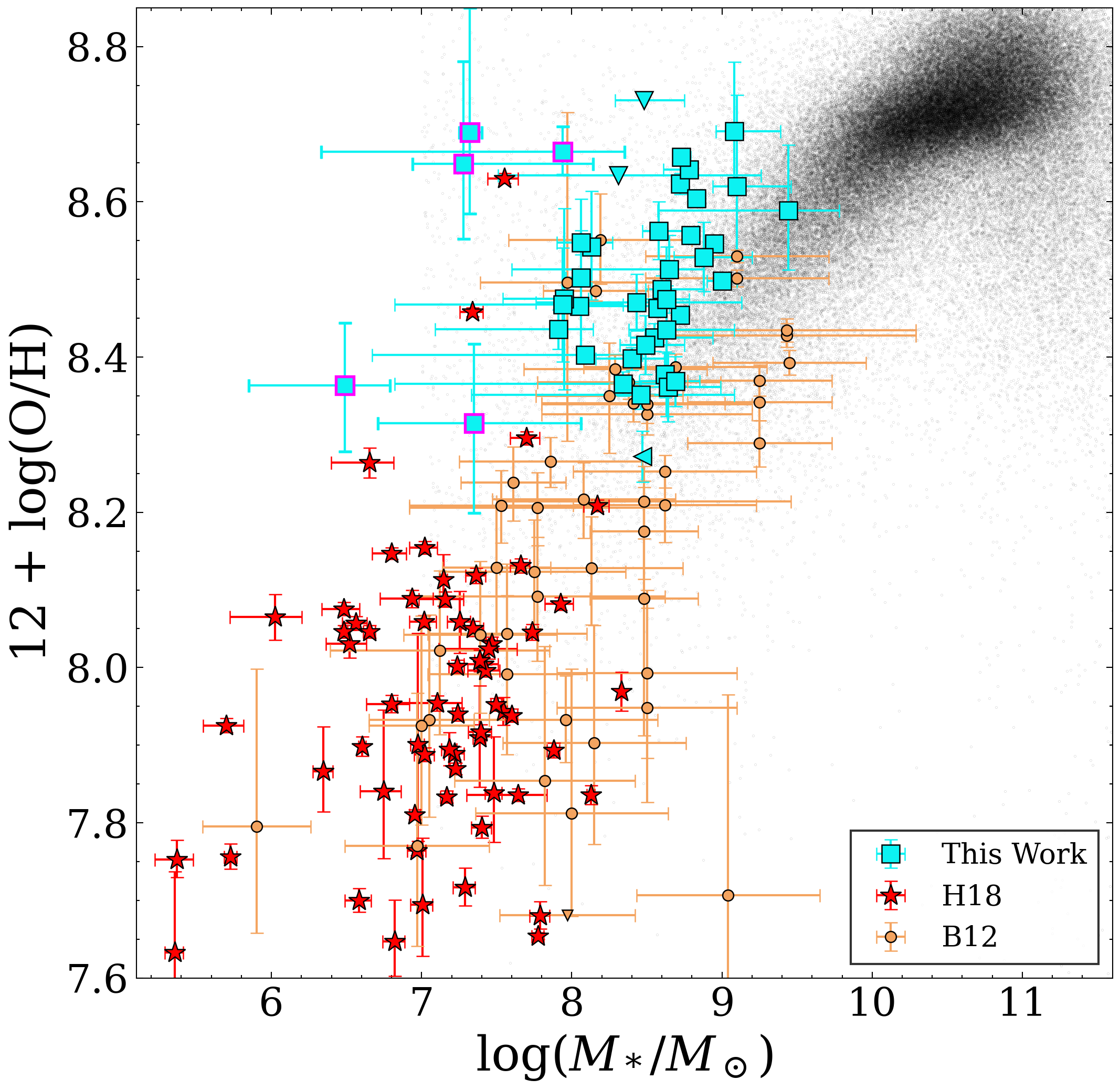}
\caption{Stellar masses and Curti N2 metallicities for our SN Ia host galaxy sample (blue squares) compared to the low-L galaxy samples of \citet[][yellow circles]{berg12} and \citet[][red stars]{hsyu18}. SDSS DR8 galaxies are shown in black. 1$\sigma$ upper limits in metallicity are shown as downward-facing triangles and upper limits in mass are shown as leftward-facing triangles. The five galaxies highlighted with a magenta outline indicate the five very low-mass galaxies we identify as having particularly high metallicities for their masses.}
\label{fig:mass_metallicity}
\end{figure}

Figure~\ref{fig:mass_metallicity} shows the distribution of our sample and the comparison samples in stellar mass and metallicity. As has been found for SN Ia host galaxies in the past \citep[e.g.,][]{childress13}, our galaxies do not clearly stand out from the other galaxy samples. They largely fall on the low-M, low-metallicity tail of the distribution of the SDSS sample and overlap with the higher metallicity galaxies from the B12 sample. The H18 sample is largely lower in mass and metallicity than our sample, but this is unsurprising, given that it was specifically selected to be low-metallicity. 

However, five of the six lowest mass galaxies in our sample have particularly high metallicities for their masses, and stand out clearly from the rest of our sample and the comparison samples. These galaxies are: ESO 357-G005 (ASASSN-15pz), GALEXASC J063224.91-713403.9 (OGLE16dha), GALEXASC J010647.95-465904.1 (ASASSN-14lw), UGCA 430 (ASASSN-16jf), and the uncatalogued host of ASASSN-15fy. We highlight these five galaxies using magenta outlines in Figure~\ref{fig:mass_metallicity}.

To test whether these galaxies truly have abnormally high metallicities when compared to the rest of our sample and the comparison samples, we fit all of the low-L samples with a linear equation of the form:

\begin{equation}
\begin{split}
12+\log(O/H) = a(\log(M_\star/M_\odot) - \mu_M) + b
\end{split}
\end{equation}

where $a$ is the slope, $b$ is the intercept, and $\mu_M$ is the mean log($M_\star$) of the sample being fit. We pivot the relations on the mean in order to make the uncertainties on the $a$ and $b$ parameters essentially uncorrelated. To robustly measure the fit and estimate the uncertainties on the $a$ and $b$ parameters, we use bootstrap resampling with 15,000 iterations. We find $(a,b)_{us}$ = $(0.60^{+0.37}_{-0.15},8.34^{+0.05}_{-0.13})$ for our sample, where the best-fit value is the median of our bootstrapped values and the errors are 1$\sigma$ uncertainties. For the B12, H1, and SDSS samples, we find $(a,b)_{B12}$ = $(0.36^{+0.09}_{-0.06},8.21^{+0.04}_{-0.04})$, $(a,b)_{H18}$ = $(0.38^{+0.15}_{-0.09},7.88^{+0.05}_{-0.07})$, and $(a,b)_{SDSS}$ = $(0.39^{+0.01}_{-0.02},8.53^{+0.01}_{-0.01})$ respectively. For the SDSS fit, we only include those galaxies with $\log{(M_{\star}/M_\odot)} \leq 9.7$ to exclude the galaxies with masses above the scale where the slope of the relation becomes shallower.

\begin{figure}
\centering
\includegraphics[width=0.99\linewidth]{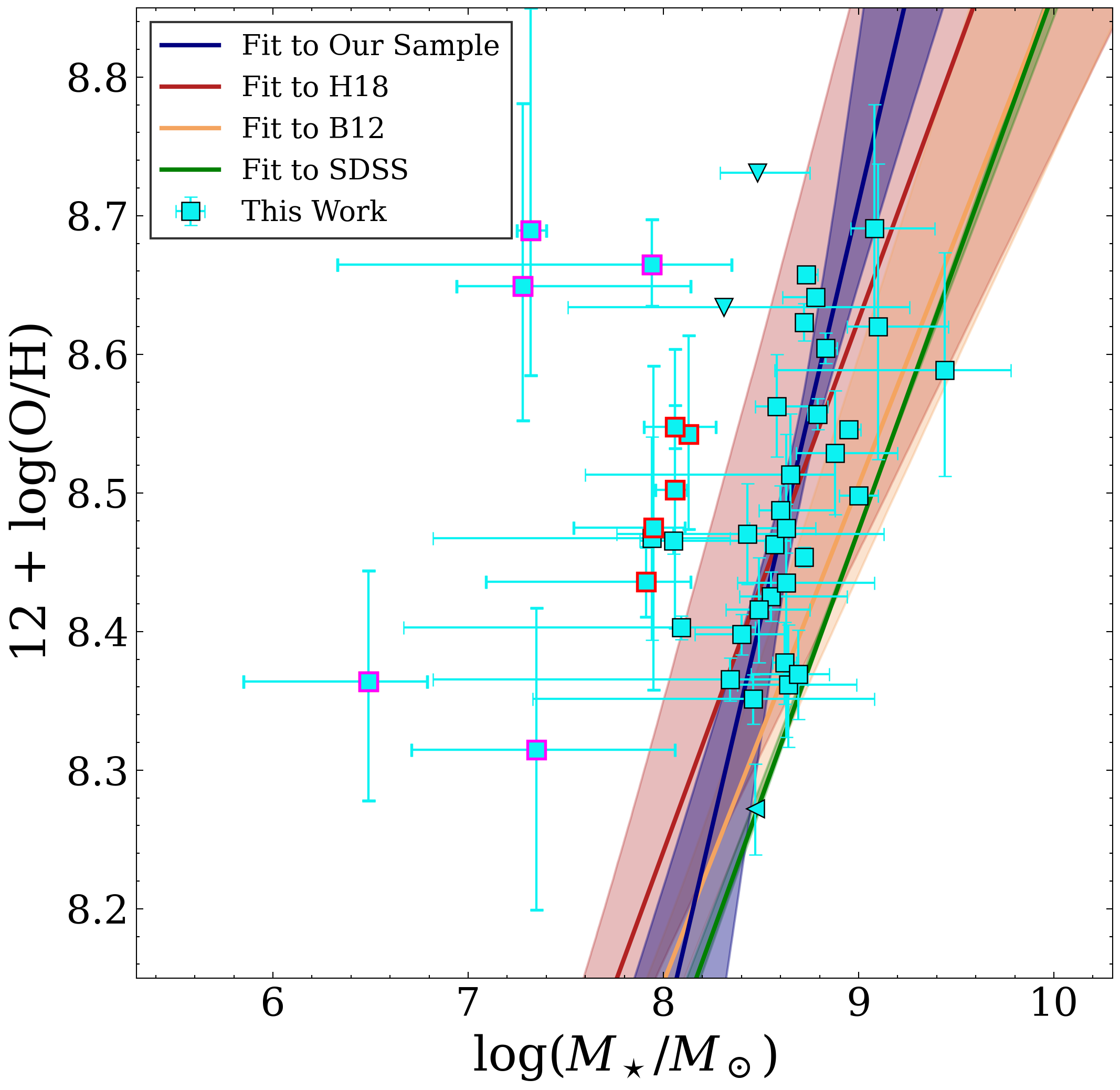}
\caption{Stellar masses and. Curti N2 metallicities for our SN Ia host galaxy sample (blue squares) compared to linear M-Z fits to our sample (navy), the B12 sample (yellow), the H18 sample (red), and the SDSS sample (green). The shaded regions in matching colors show the $1\sigma$ range for each fit. Note the change in scales compared to Figure~\ref{fig:mass_metallicity}, which we have adjusted to more clearly focus on our sample. The five data points noted in Figure~\ref{fig:mass_metallicity} as having high metallicities are highlighted in magenta, and five remaining galaxies that are inconsistent with any of the fits at the 1$\sigma$ level are highlighted in red.}
\label{fig:mz_fits}
\end{figure}

In Figure~\ref{fig:mz_fits} we show our sample as well as the best-fit lines and 1$\sigma$ uncertainties for all four galaxy samples. All three fits are generally quite consistent through the mass range shown, particularly those for the B12 and H18 samples. Four of the five galaxies we identified as having high metallicities are inconsistent with all four fits at the 1$\sigma$ level. The fifth is mildly consistent with only the H18 fit but inconsistent with the others. We also identify five additional galaxies which are inconsistent with all four fits at the $1\sigma$ level: GALEXASC J090013.19-133803.5 (ASASSN-16oz), GALEXASC J104848.62-201544.1 (ASASSN-16dn), GALEXASC J215327.92-342420.8 (ASASSN-16hw), the uncatalogued host of CSS141227:132741-111252, and the uncatalogued host of Gaia16alq. These five galaxies and the five identified above are 10 of the 14 lowest-mass galaxies in our sample. Of the nine galaxies with $\log{(M_\star/M_\odot)}<8$ in our sample, seven are inconsistent with any of the fits, and all nine are inconsistent with every fit except for the H18 fit. The H18 fit is more uncertain at the metallicities of our sample because the H18 galaxies lie at lower metallicities, and the uncertainties on the fit increase further from the locus of the data, making it the least reliable correlation for the metallicities spanned by our sample. We thus conclude that there is significant evidence from our sample that the lowest mass SN Ia host galaxies are more metal-rich than typical low-mass galaxies.

\begin{figure}
\centering
\includegraphics[width=0.99\linewidth]{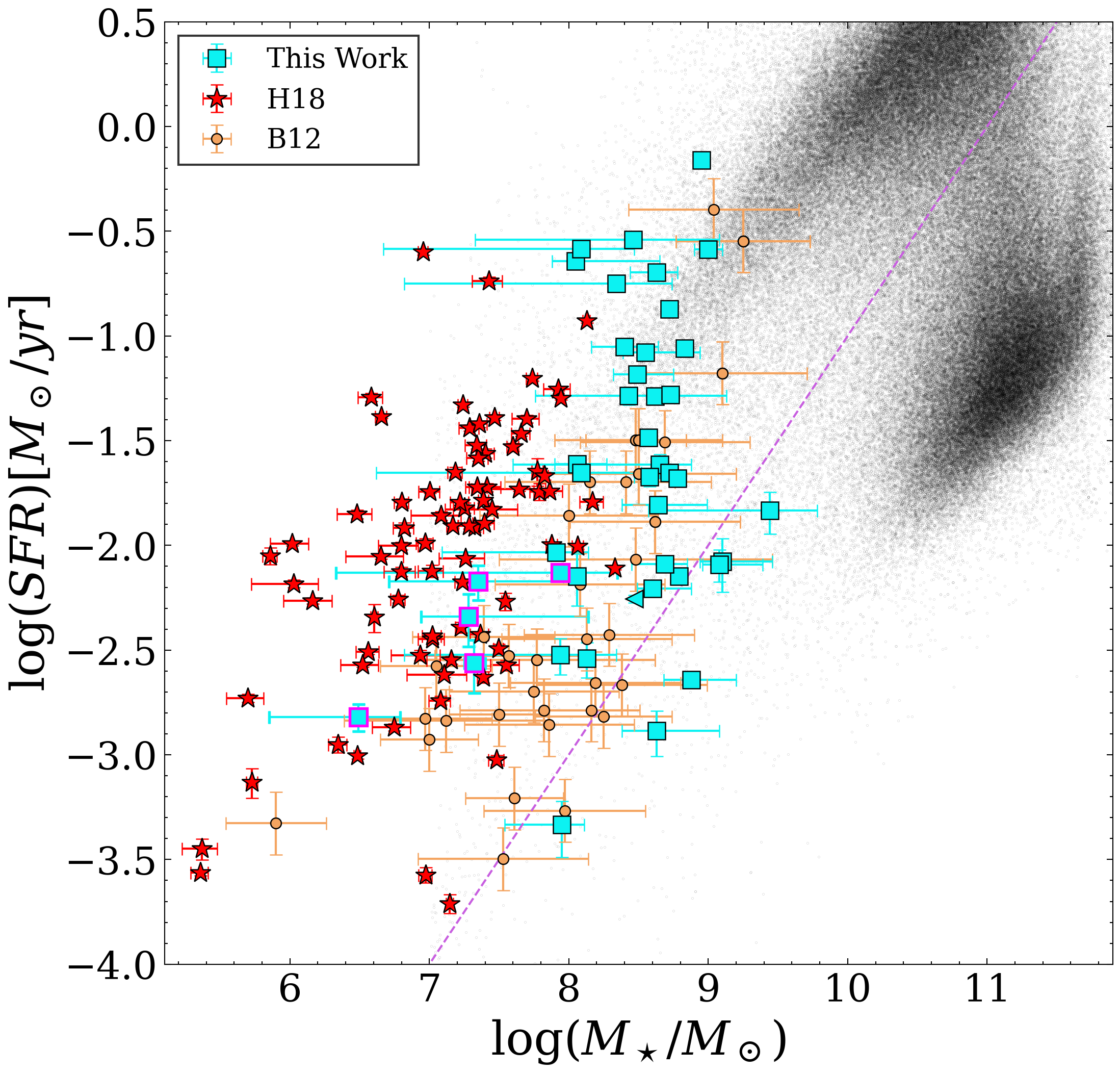}
\caption{The stellar masses and SFRs of our sample compared to our comparison samples. Colors and symbols match those of Figure~\ref{fig:mass_metallicity}, with the five points highlighted with magenta outlines indicating the five galaxies with notably high metallicities for their masses. The magenta dashed line shows the division between actively star-forming galaxies and passive galaxies as defined in \citet{brown19}, corresponding to log(sSFR)$<-11$.}
\label{fig:mass_sfr}
\end{figure}

Figure~\ref{fig:mass_sfr} shows the stellar masses of our sample and the comparison samples compared to their current SFRs. The magenta line in the Figure corresponds to log(sSFR)$<-11$, and is the division used by \citet{brown19} to separate actively star-forming and passive galaxies. As expected based on their locations in the diagnostic diagrams of Figure~\ref{fig:line_diagnostics}, the majority of our galaxies lie above this dividing line, implying that they are actively star-forming. The six galaxies below the line are all very close to the dividing line, and of these only four are inconsistent with being above the line given the uncertainties on their masses and SFRs. Our sample does not clearly stand out from any of our comparison samples, and is quite similar to the B12 sample in particular. The five galaxies noted as having high metallicities for their masses are again highlighted in magenta in the figure. While they do have some of the lower SFRs of the galaxies in our sample, they do not stand out from the other galaxies. We conclude that there is nothing atypical about the SFRs our sample given their masses. 

\begin{figure}
\centering
\includegraphics[width=0.99\linewidth]{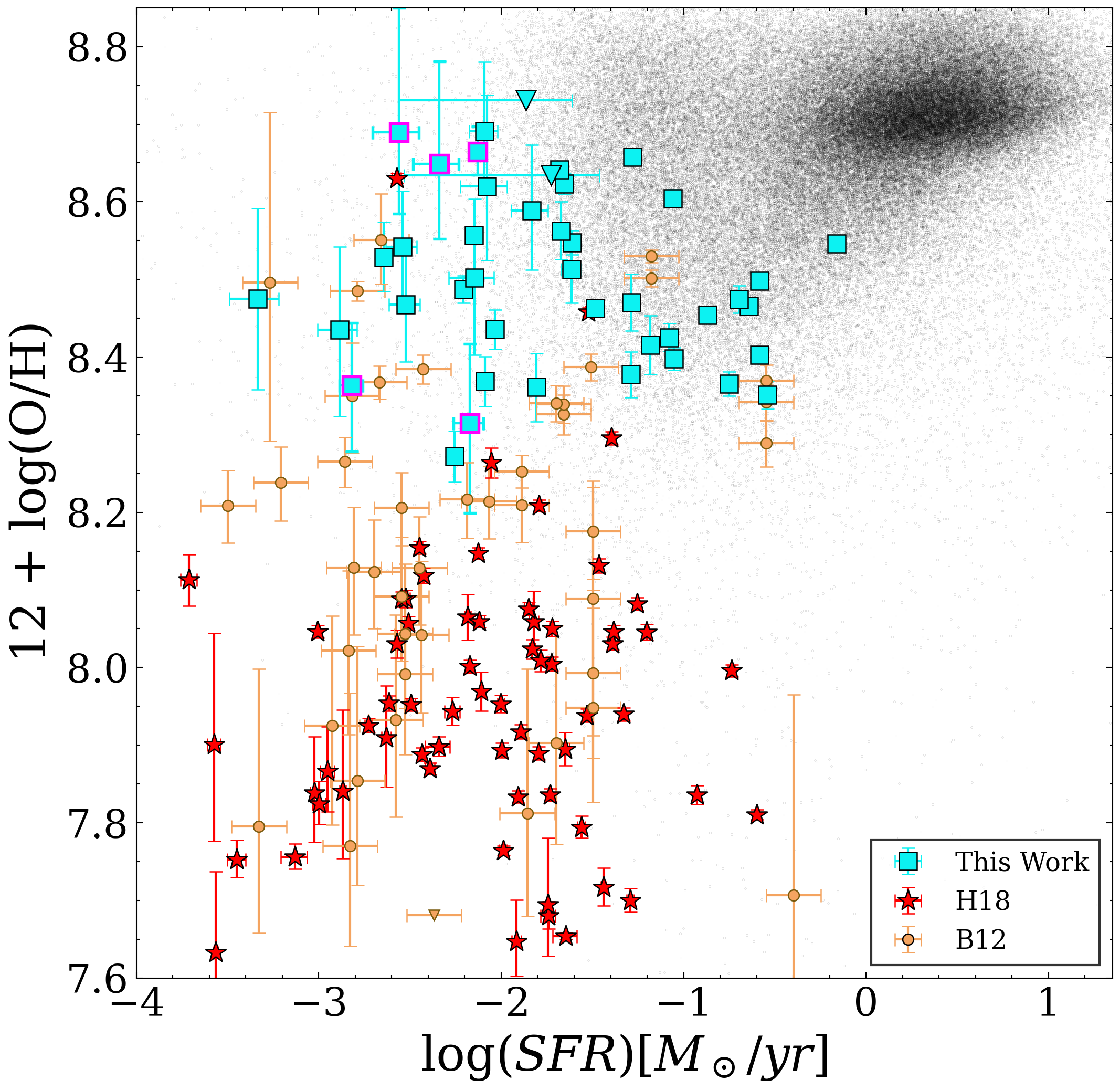}
\caption{The SFRs and Curti N2 metallicities for our sample compared to our galaxy comparison samples. Colors and symbols match those of Figure~\ref{fig:mass_metallicity}, with the 5 points highlighted with magenta outlines indicating the 5 galaxies with notably high metallicities for their masses. 1$\sigma$ upper limits in metallicity are shown as downward-facing triangles.}
\label{fig:sfr_metallicity}
\end{figure}

In Figure~\ref{fig:sfr_metallicity} we show the SFRs and metallicities of our sample. Our galaxies tend to have higher metallicities than the galaxies with similar SFRs in our low-L comparison samples, but are generally consistent with the outer edge of the SDSS sample. There is little overlap between our sample and the H18 sample, while the B12 sample is fairly evenly spread throughout the region spanned by our sample and the H18 sample. There is no clear trend in metallicity with SFR for the three low-L samples shown here, with the metallicity spread for each sample being fairly flat over a wide range of SFRs. The five high-metallicity galaxies noted in magenta do not stand out from the rest of the galaxies in our sample in the SFR-metallicity plane.

\begin{figure}
\centering
\includegraphics[width=0.99\linewidth]{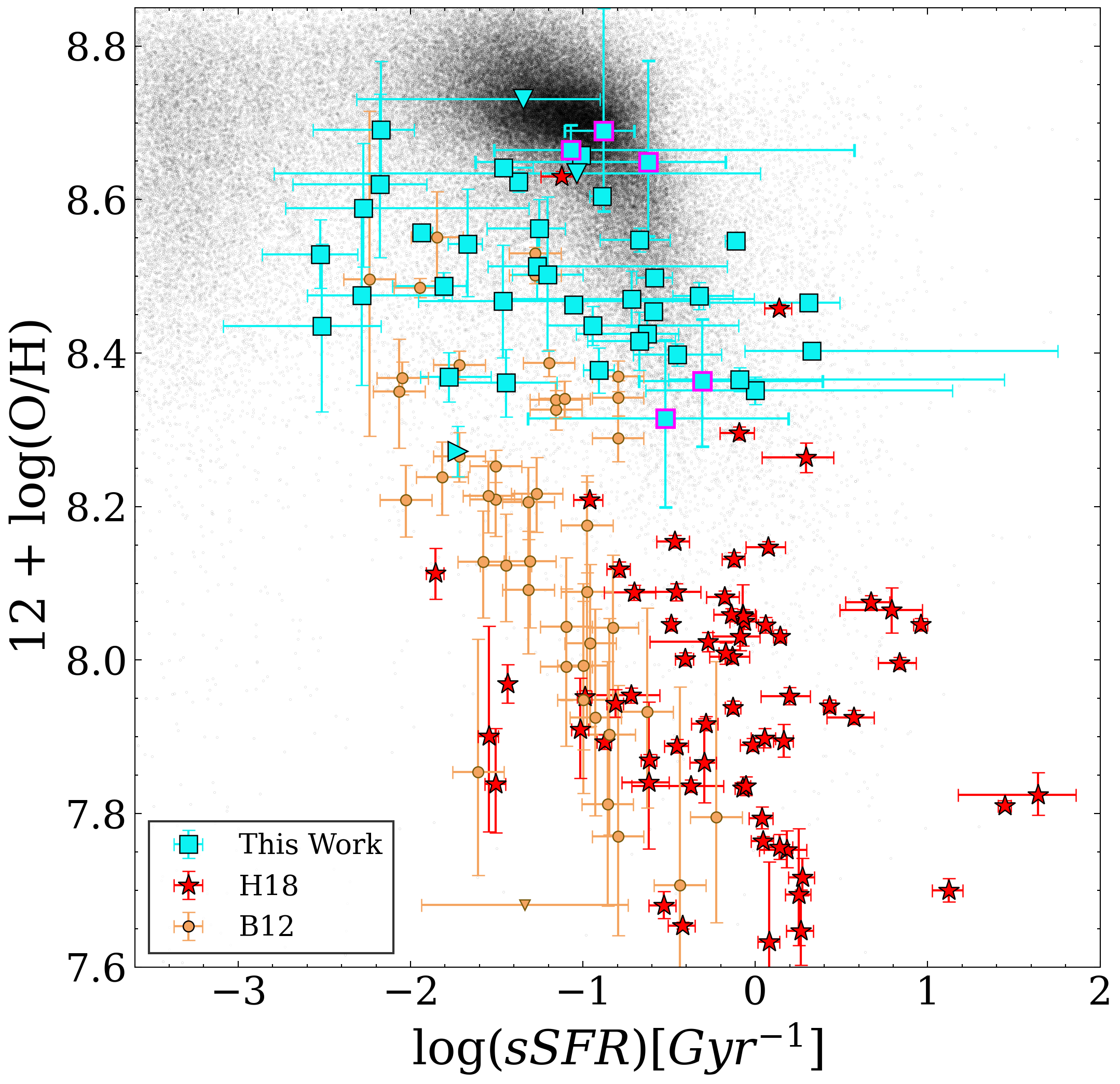}
\caption{The sSFRs and Curti N2 metallicities of our sample compared to the comparison samples. Colors and symbols match those of Figure~\ref{fig:mass_metallicity}, with the 5 points highlighted with magenta outlines indicating the five galaxies with notably high metallicities for their masses. 1$\sigma$ upper limits in metallicity are shown as downward-facing triangles and lower limits in sSFR are shown as right-facing triangles.}
\label{fig:ssfr_metallicity}
\end{figure}

In Figure~\ref{fig:ssfr_metallicity} we examine the sSFRs of our sample and comparison samples compared to their metallicities. We again see that our sample is higher in metallicity compared to the other low-L samples, with even less overlap between our sample and the B12 and H18 samples than in the space of SFR and metallicity. Our sample is again generally consistent with the SDSS DR8 sample, however. We again highlight the five galaxies identified as having high metallicities for their masses, and note that they do not seem to have atypical metallicities given their sSFRs, and three of them fall in the area of the Figure most densely populated by SDSS galaxies. Where our galaxies show no obvious trend in metallicity with increasing sSFR, as was the case with SFR, this does not appear to be true for the other low-L samples. The B12 sample has a clear trend with the metallicity decreasing as sSFR increases. While the H18 sample does not have as clear of a trend overall, we note that the highest metallicity galaxies in the H18 sample tend to have higher sSFRs. 

For both SFR and sSFR, our sample does not appear to be particularly metal-rich or metal-poor compared to the general population of galaxies from SDSS. In comparison to similar low-L samples in B12 and H18, however, our sample is more enriched for a given SFR/sSFR. Neither SFR nor sSFR appears to have a clear correlation with metallicity in our sample, despite there being a trend with mass.



\section{Discussion}
\label{sec:disc}

Our spectroscopic examination of 44 low-luminosity host galaxies of SNe Ia indicates that this population of galaxies is generally similar to typical galaxies in metallicity, SFR, sSFR, and stellar mass. In particular, the SFRs and sSFRs of our sample appear to be quite normal given the galaxies' masses and metallicities. These results generally support the conclusions of previous examinations of SN Ia host galaxies \citep[e.g.,][]{childress13}, which have found that SN Ia hosts generally have metallicites similar to galaxies of the same masses and SFRs.

We do, however, find that 10 of the 14 least massive galaxies in our sample are inconsistent at a $1\sigma$ level with the mass-metallicity relations defined by our comparison galaxy samples (which are generally consistent with one another, and with the M-Z relation defined by our entire sample). This includes five galaxies that have notably high metallicities for their masses. Further, all nine of the galaxies with log($M_\star/M_\odot$)$<8$ are inconsistent with the mass-metallicity fits for all but the least certain fit at the masses spanned by our sample. We conclude that there is thus some evidence that the lowest-mass, lowest-luminosity host galaxies of SNe Ia are more metal-rich than general galaxies.

This conclusion is not without caveats, however. First, while we have taken steps to ensure that our sample of longslit and IFU spectra are comparable to one another and to the comparison samples we use, in some cases our longslit spectra do not capture all of the host galaxy light. This would result in some inaccuracies of the properties we derive for our galaxies, such as SFR and metallicity, that are based on spectroscopic line fluxes. As mentioned in Sections~\ref{sec:sample_details} and \ref{sec:phys_props}, we have checked our results for consistency using some galaxies with multiple observations in our sample and by comparing our spectroscopically derived SFRs to the photometrically derived ones from \citet{brown19}, and find that our results are consistent. We thus believe effects of slit losses on our conclusions are minimal. 

A more substantial caveat with the conclusion that low-mass galaxies in our sample have higher metallicities is due to the size of our sample. Though our sample contains more low-luminosity galaxies than previous SN Ia host samples, it only includes nine galaxies with log($M_\star/M_\odot$)$<8$, where we see deviations from the mass-metallicity relations. The outlier galaxies are also largely discrepant at only the $1\sigma$ level. Thus, while we do believe there is evidence that these galaxies do have high metallicities for their masses, we cannot rule out that the effect we see is the result of small number statistics. Larger samples of low-luminosity and low-mass SN Ia host galaxy samples, particularly those with log($M_\star/M_\odot$)$<8$, are needed to determine if these conclusions are robustly supported. 

If our results are indeed true, this would have interesting implications for both SN studies and host galaxy studies. Studies of SN host galaxies often use photometry to infer physical properties of the galaxies being studied, as photometry typically requires significantly less observational resources than spectroscopy. While methods such as photometric SED fitting can obtain fairly reliable masses and even SFRs, metallicity is typically calculated using a mass-metallicity relation, and cannot be measured directly without spectroscopy. If low-luminosity SN Ia hosts deviate from standard mass-metallicity relations, the inferred metallicities for these galaxies would be too low. It is thus important to determine whether the effect seen in our study is supported by additional data, and correct metallicity studies of SN Ia host galaxies if so.

These findings may impact our understanding of SNe Ia as well. It has previously been seen that higher host galaxy metallicity results in brighter, faster declining, and redder SNe Ia \citep[e.g.,][]{howell09,pan13}. While these have typically also been more massive and older hosts in previous samples, if SNe in our sample look similar, it would be a clear sign that metallicity is the primary attribute that regulates these SN properties. Alternatively, if SNe in these hosts resemble those in other low-mass, young galaxies, it would imply that metallicity does not play a strong role in determining these aspects of SNe Ia. Examination of the SN Ia rate and SN Ia properties in this low-luminosity population of hosts could thus have implications on our understanding of SN Ia progenitor systems and how the metallicity of the progenitor affects the SN explosion. This in turn could also affect cosmological studies based on SNe Ia.

Apart from the interesting discovery of potential low-mass outliers in host galaxy metallicity, this work provides the spectroscopic observations needed to perform studies on specific SN Ia rates and SN property-host property correlations, as described above. In future work we will combine this sample with the much larger sample of more luminous and massive ASAS-SN SN Ia host galaxies that have archival spectroscopic observations. This will allow for a comprehensive study of how host environment affects SN Ia progenitors and explosions in the low-redshift Universe, which will include a substantial fraction of low-luminosity hosts for the first time.

\acknowledgments

The authors thank P. Senchyna, and G. Rudie for their assistance with finding comparison samples.

Support for T.W.-S.H. was provided by NASA through the NASA Hubble Fellowship grant HST-HF2-51458.001-A awarded by the Space Telescope Science Institute (STScI), which is operated by the Association of Universities for Research in Astronomy, Inc., for NASA, under contract NAS5-26555. L.G. acknowledges financial support from the Spanish Ministerio de Ciencia e Innovaci\'on (MCIN), the Agencia Estatal de Investigaci\'on (AEI) 10.13039/501100011033, and the European Social Fund (ESF) "Investing in your future" under the 2019 Ram\'on y Cajal program RYC2019-027683-I and the PID2020-115253GA-I00 HOSTFLOWS project, from Centro Superior de Investigaciones Cient\'ificas (CSIC) under the PIE project 20215AT016, and the program Unidad de Excelencia Mar\'ia de Maeztu CEX2020-001058-M. P.J.V. is supported by the National Science Foundation Graduate Research Fellowship Program Under Grant No. DGE-1343012. C.S.K. and K.Z.S. are supported by NSF grants AST-1814440 and AST1908570. H.K. was funded by the Academy of Finland projects 324504 and 328898. J.D.L. acknowledges support from a UK Research and Innovation Future Leaders Fellowship (MR/T020784/1). Support for J.L.P. is provided in part by ANID through the FONDECYT regular grant 1191038 and through the Millennium Science Initiative grant ICN12009, awarded to The Millennium Institute of Astrophysics, MAS. 

The LBT is an international collaboration among institutions in the United States, Italy and Germany. LBT Corporation partners are: The Ohio State University, and The Research Corporation, on behalf of The University of Notre Dame, University of Minnesota and University of Virginia; The University of Arizona on behalf of the Arizona university system; Istituto Nazionale di Astrofisica, Italy; LBT Beteiligungsgesellschaft, Germany, representing the Max-Planck Society, the Astrophysical Institute Potsdam, and Heidelberg University.

\software{IRAF (Tody 1986, Tody 1993)}

\bibliography{bibliography.bib}{}
\bibliographystyle{aasjournal}



\end{document}